\documentstyle[12pt,epsf]{article}
\def\ga{\mathrel{\raise.3ex\hbox{$>$\kern-.75em\lower1ex\hbox{$\sim$}}}}
\def\la{\mathrel{\raise.3ex\hbox{$<$\kern-.75em\lower1ex\hbox{$\sim$}}}}
\def\gev{{\rm \, Ge\kern-0.125em V}}
\def\tev{{\rm \, Te\kern-0.125em V}}
\def\beq{\begin{equation}}
\def\eeq{\end{equation}}

\def\ss{\scriptscriptstyle}

\def\mpar{m_{\ss\|}^2}
\def\mpl{M_{\rm Pl}}

\begin{document}
\begin{titlepage}
\pagestyle{empty}
\baselineskip=21pt
\rightline{hep-ph/9510308}
\rightline{UMN--TH--1411/95}
\rightline{TPI--MINN--95/29}
\rightline{UCSBTH--95--31}
\rightline{October 1995}
\vskip1.25in
\begin{center}
{\large{\bf
New Constraints on Superpartner Masses}}
\end{center}
\begin{center}
\vskip 0.5in
{Toby Falk,$^1$ Keith A.~Olive,$^1$  Leszek Roszkowski,$^1$
and Mark Srednicki$^2$
}\\
\vskip 0.25in
{\it
$^1${School of Physics and Astronomy,
University of Minnesota, Minneapolis, MN 55455, USA}\\
$^2${Department of Physics,
University of California, Santa Barbara, CA 93106, USA}\\}
\vskip 0.5in
{\bf Abstract}
\end{center}
\baselineskip=18pt \noindent
We consider the Minimal Supersymmetric Standard Model (MSSM)
without imposing relations on the superpartner masses that arise in
grand unified theories.  Given an arbitrary pattern of superpartner masses
(consistent with experimental constraints), it may happen that the scalar
potential is actually unstable, even though all scalar masses-squared are
positive at the weak scale $M_W$.  This is most likely to happen if the
running mass-squared in a ``flat'' direction in field space becomes negative
at some scale $Q_0$ which is well below the GUT scale.  In this case, either
this pattern of masses is ruled out, or there must be new physics
(beyond the MSSM) at or below the scale $Q_0$.
\end{titlepage}
\baselineskip=18pt
The appearance of supersymmetry at the electroweak scale is widely
regarded as a likely possibility.  The main problem in analyzing
supersymmetric extensions of the Standard Model is the huge number
of new parameters.  Even in the Minimal Supersymmetric Standard Model (MSSM),
we must specify mass matrices for the superpartners; these masses
softly break supersymmetry, and are arbitrary unless we make some
assumptions about their ultimate origin.  Of course, some of these
new parameters are restricted by experimental constraints,
in particular, by the absence of flavor-changing neutral currents.
Still, it is always necessary to make some sort of simplifying
assumptions in order to deal with the remaining complexity.

One of the best motivated assumptions is that of conventional grand
unification, with the further assumption that all squark and slepton
masses are equal at the grand unified (GUT) scale \cite{ws,ikkt}.
However, there are a number of
reasons to doubt these assumptions.
First of all, even in a GUT, there is no reason
for the squark and slepton masses of different generations
to be equal. In addition, superstring models
need not involve conventional grand unification, and can in fact
give very different mass patterns \cite{m1}.  The low energy content of the
theory is often enlarged, with extra U(1) gauge bosons and extra
scalars (moduli fields), and of course their superpartners.
Also, it may be that supersymmetry is broken by extra strong
interactions (technicolor) at relatively low energies, leading
to still more complicated scenarios \cite{m2}.

Our purpose here is to point out that certain mass patterns for the
superpartners cannot arise unless there is new physics (beyond the MSSM)
well below GUT scale.  To see this, consider using the
renormalization group equations to run the superpartner masses-squared
up from the weak scale.  It may happen that some squark or slepton
mass-squared parameters become negative at some scale $Q_0$ which is
well below the GUT scale, apparently leading to large
and disastrous vacuum expectation values for the corresponding fields.
If this is the case, clearly this region of parameter space is ruled out,
or we made a mistake in trusting the renormalization group equations
up to the scale $Q_0$.  This would imply the existence of new physics
at or below the scale $Q_0$.
However, there is an important caveat:
the computed VEVs are not trustworthy if they are much
less than the renormalization scale $Q>Q_0$ which we choose to employ.
This is because there are large
logarithms in the loop corrections which cannot be absorbed into
the running parameters.  If, on the other hand, the VEVs turn out to be
roughly equal to (or greater than) $Q$, then disastrous breaking of color
and/or electric charge can be expected to occur.

Constraints on the parameter space (most notably on the
soft supersymmetry breaking trilinear couplings $A$) arising from the
existence of charge and/or color breaking minima have been previously
considered \cite{klnq}.
As far as we know, however, previous authors have considered
only directions in field space involving VEVs of Higgs fields,
whereas the specific example we consider involves squark fields only.

Our analysis will use methods used previously to study vacuum
stability in the MSSM (e.g., \cite{quiros} and references therein).
There are two cases which can be considered separately.
In the first, all the VEVs appear in directions in field space in which the
tree-level potential has nonvanishing quartic terms.  (There may
also be cubic terms, but these will not be qualitatively important,
and we will neglect them for simplicity.)  Schematically, we have
\begin{equation}
V=\frac12 m^2(Q)\phi^2 + \frac14 \lambda(Q)\phi^4,
\label{pot}
\end{equation}
The scalar masses, $m^2(Q)$, are affected by their gauge and Yukawa
interactions. For a given set of masses at the weak scale, interactions with
gauginos drive the scalar masses down as one moves to higher energy scales
whereas their Yukawa interactions tend to drive them up \cite{ikkt}.
The  Yukawa interactions are important only for stops (and perhaps for
sbottoms and staus if $\tan \beta$ is large). Depending on the particular
pattern of sfermion masses at the weak scale, it may happen that
the interactions
with gauginos in fact drives $m^2(Q)$ negative, in which case the scalar
potential has a minimum at $\phi = v(Q)=[-m^2(Q)/\lambda(Q)]^{1/2}$.
We will assume that $m^2(Q)$ is negative when $Q$ is greater than some
particular scale $Q_0$.

The presence of the minimum at $\phi=v(Q)$ can be trusted only if it is stable
with respect to radiative corrections; in general, at the one-loop level
these are of the form $\Delta V \sim v(Q)^4 \ln [v(Q)^2/Q^2]$ \cite{cw}.
Hence, the minimum at $\phi=v(Q)$ is trustworthy if $v(Q)\simeq Q$,
where the one-loop corrections can safely be assumed to be small.
Thus, the best value of $Q$ to use is the one which yields $v(Q)=Q$.
If $Q_0$ is large, then it can happen that this equation has no solution.
This indicates that the true, renormalization-group invariant value
of the VEV is zero.  In general, however, we usually find that there
is, in fact, a solution to $v(Q)=Q$, and this means that either
(1) the pattern of superpartner masses which results in a nonzero VEV
is ruled out, or (2) new physics must appear at or below the scale $Q$,
new physics which somehow forces the VEV back to zero.

The analysis outlined above is somewhat complicated, since there are
many directions in field space to check.  Luckily, we find that the strongest
constraints arise in the second case: directions
in field space for which $\lambda(Q)=0$ for all $Q$.
This case is the simplest to analyze, and
the MSSM has, in fact, many such ``flat'' directions.
Let us denote by $\mpar(Q)$ the mass-squared
along one of these flat directions.  Then if a negative $\mpar(Q)$ is found
for $Q$ greater than some value $Q_0$,
the potential is unbounded below for large field values
in the flat direction.  This unboundedness might still be cured, but only by
nonrenormalizable terms in the scalar potential, such as
$V_{\rm NR}(\phi) \sim M^{4-n}\phi^n$, where $n>4$, and $M$ is a new mass
scale (e.g., the Planck mass or an intermediate scale)
corresponding to new physics.
In this case, the field will acquire a VEV
\beq
v(Q) \sim M^{(n-4)/(n-2)}[-{\mpar}(Q)]^{1/(n-2)}
\eeq
and for $M>Q_0$, there will usually be a solution of $v(Q)=Q$.

There are many flat directions which in
principle must be checked.  We concentrate on a particular example,
the direction $\tilde u_R^{\ss r} = \tilde s_R^{\ss g} = \tilde
b_R^{\ss b} \equiv v(Q)$ \cite{ad}.
We expect this direction to provide particularly strong bounds, as it
involves only squarks (whose masses run faster than the sleptons
due to their strong coupling to gluinos) and it does not involve the
stop mass (whose running is slowed by Yukawa interactions).
For this case we have
$\mpar = m_{\tilde u_{\ss R}}^2 + m_{\tilde s_{\ss R}}^2
+ m_{\tilde b_{\ss R}}^2$, whose renormalization group equation is
\begin{equation}
Q{{\rm d}\mpar\over{\rm d}Q} = {1\over 8\pi^2}
\left[-16g_3^2 M_3^2 - {8\over 3} g_1^2 M_1^2
+ 2 h_b^2 \left(m^2_{\tilde q_{\ss L}} + m^2_{\tilde b_{\ss R}}
+ m_{H_1}^2 + A_b^2 \right)\right],
\label{mparrge}
\end{equation}
where $g_1$ is the standard model $U(1)$ coupling,
$M_i$ is a gaugino mass, $h_b$ is the bottom-quark Yukawa coupling,
and $A_b$ is the bottom-quark trilinear mixing parameter; we have neglected
terms involving first and second generation Yukawa couplings\footnote{
Note that for this flat direction we have a cancellation of the
U(1) $D$-terms which must be included for generic patterns of scalar
masses.  These contributions are absent in GUT models, where one assumes that
the scalar masses are all equal at some scale \cite{ikkt,mv}.}.
The full set of RGEs can be found, for example, in \cite{ikkt,dn,mv}.
If $\tan\beta$ (the ratio of the two Higgs VEVs) is not too large,
then $h_b$ is small, and the term proportional to $h_b^2$ has a
nearly negligible effect.  We see that, as $Q$ increases, $\mpar$
decreases, and may become negative.

Eq.(\ref{mparrge}) can be solved in closed form by making use of
the one-loop relations $M_i(Q)\propto g_i^2(Q)$ and
$g_i^2(Q_1) = g_i^2(Q_2)/[1 - b_i g_i^2(Q_2)\ln(Q_1/Q_2)/(8\pi^2)]$
with $b_3=-3$ and $b_1=+11$.  Assuming that
the physical (propagator pole) squark masses entering
$\mpar$ are less than the gluino mass $M_3$ and the bino mass
$M_1$, then the solution of eq.(\ref{mparrge}) is
\begin{eqnarray}
\mpar(Q)=\mpar
&-& {2\over\pi^2} g_3^2(M_3)M_3^2 \ln(Q/M_3)
\left\{{1+3g_3^2(M_3)\ln(Q/M_3)/(16\pi^2) \over
\left[1+3g_3^2(M_3)\ln(Q/M_3)/(8\pi^2)\right]^2} \right\}
\nonumber \\
&-& {1\over3\pi^2} g_1^2(M_1)M_1^2 \ln(Q/M_1)
\left\{{1-11g_1^2(M_1)\ln(Q/M_1)/(16\pi^2) \over
\left[1-11g_1^2(M_1)\ln(Q/M_1)/(8\pi^2)\right]^2} \right\}
\label{rgesoln}
\end{eqnarray}
where all masses on the right-hand side are physical (propagator pole)
masses.  The only undetermined factors are $g_3^2(M_3)$ and
$g_1^2(M_1)$, which  depend on the full spectrum of superpartner masses
(via threshold effects).

We have solved the full set of RGEs numerically, and
 in fig.~(1), we show contours
of constant $v(Q)=Q$ in the $m_{\ss\|}/\sqrt3$--$M_3$ plane;
note that the $m_{\ss\|}/\sqrt3$ is the root-mean-square average of the
three relevant squark masses.  We have taken
$\alpha_3(M_Z)=0.12$, $A_{b,t}=0$,
the higgsino mass parameter $\mu=1000\,$GeV,
$\tan\beta=3$, and the pseudoscalar Higgs mass $m_A=400\gev$.
We have actually run the RGE's for several sets of parameter
choices, and we find negligible variation with respect to $A,\mu$,
and $\tan \beta$, if $\tan \beta \la 10$.
Also, we have taken $M_1=M_2=M_3$ as an illustrative choice;
our results are not at all sensitive to the value of $M_2$,
and sensitive to the value of $M_1$ only if $M_1 \gg M_3$.
Furthermore, we have taken
$V_{\rm NR}(\phi)={1\over6}M^{-2}\phi^6$ with $M=\mpl=1.2\times 10^{19}\,$GeV.
The value of $v(Q)$ which labels the contours depends on this choice,
roughly as $v(Q)\propto M^{1/2}$, so that lowering $M$ implies new physics
at a lower scale.
However, the location of the uppermost contour does not depend on $M$.
Above this contour, there is no solution of $v(Q)=Q$,
and so there is no unphysical
minimum of the scalar potential; these values of $m_{\ss\|}$ and $M_3$,
given by the approximate formula
\begin{equation}
m_{\ss\|}/\sqrt3 \ga 0.7 M_3
\label{contour}
\end{equation}
for $M_3 < 2000\,$GeV, are allowed.
Below this contour, there is a solution
of $v(Q)=Q$, and so either these values of
$m_{\ss\|}$ and $M_3$ are
ruled out, or new physics must appear at or below the scale $Q$
(which depends on the choice of $V_{\rm NR}$).
The most exciting possibility is the eventually experimental
discovery of squarks and gluinos with masses which do not
satisfy eq.~(\ref{contour}), as this would predict new physics
(beyond the MSSM) at or below the corresponding value of $Q$
shown in fig.~(1). Note that based on our constraint, we do not expect
to find squark masses with $m_{\ss\|}$ lower than our lowest contour
labeled $10^3$ GeV, as we do not expect new physics below this scale.
In addition, for $M< \mpl$ our constraints on $m_{\ss\|}$
are strengthened.  Similar
constraints can also be found for other combinations of squarks and
sleptons along other flat directions.

Our constraint involves fairly large squark masses,
and these can have cosmological consequences.
We must ensure that the lightest supersymmetric particle (LSP)
has an efficient annihilation channel, so that their relic mass
density does not overclose the universe.
If this particle is a gaugino, annihilation via
squark and slepton exchange is the dominant mechanism, and
so the squarks and sleptons cannot all be too heavy \cite{os34}.
The large sfermion masses implied by our constraints lead to a
relic density of gauginos
$\Omega_{\widetilde \chi}h^2\gg 1$ unless either some combination of
squark and slepton masses are light while still satisfying the bound on
$m_{\ss\|}$ or there is significant mixing among the stops providing us with
a relatively light stop.  (This possibility requires of course that the LSP
is heavier than the top quark \cite{fmos}.)
Thus, the commonly made ansatz of taking equal squark and slepton masses
at the weak scale is cosmologically excluded unless
$m_{\widetilde \chi} > m_t$ and we have a light stop.  When
the LSP is a higgsino, for which
annihilation through intermediate sfermions is not as important, these
cosmological restrictions do not apply.
However, for a Higgsino LSP, $\Omega_{\widetilde \chi}h^2\ll 1$,
unless $m_{\widetilde \chi} \ga 1$ GeV.

To conclude, we have identified a new constraint on supersymmetric
models which do not include conventional grand unification.
The sparticle masses at the weak scale
must be such that the squark and slepton
fields do not acquire large VEVs.  Charge  and/or color breaking minima  can
occur if a squark or slepton mass-squared becomes negative at some scale $Q_0$.
Directions in field space for which the quartic term vanishes (which are in
fact
common in supersymmetric models) are particularly susceptible to the formation
of large VEVs.  In general, if the effective potential at scale $Q$ predicts a
VEV of a squark or slepton field
$v(Q)$ such that $v(Q)=Q$, then new physics must appear at or below
this value of $Q$.

\vskip 0.5in
\vbox{
\noindent{ {\bf Acknowledgments} } \\
\noindent  We would like to thank D. B\"{o}decker,
B. Campbell, S. Chaudhuri, H. Haber, A. Kovner,
S.P. Martin, M. Voloshin, and H. Weigert
for helpful discussions.
This work was supported in part by DOE grant DE--FG02--94ER--40823
and NSF grant PHY--91--16964.}

\newpage
\vskip 2in
\noindent{\bf{Figure Captions}}

\vskip.3truein

\begin{itemize}
 \item[]
\begin{enumerate}
\item[]
\begin{enumerate}

\item[Fig.~1)]Contours of constant $v(Q)=Q$ as a function of the
gluino mass $M_3$ and the root-mean-square average squark
mass $m_{\ss \|}$. We take a non-renormalizable operator of the
form $V_{NR} = {1\over6} \phi^6/\mpl^2$ to stabilize the scalar potential
at high scales.  Successive contours represent an increase by a
factor of $10$ in $v(Q)$.   Regions below a contour are forbidden unless
new physics appears below that scale. There are no solutions to $v(Q)=Q$ for
$v\ga 10^{10}\gev$.

\end{enumerate}
\end{enumerate}
\end{itemize}
\newpage

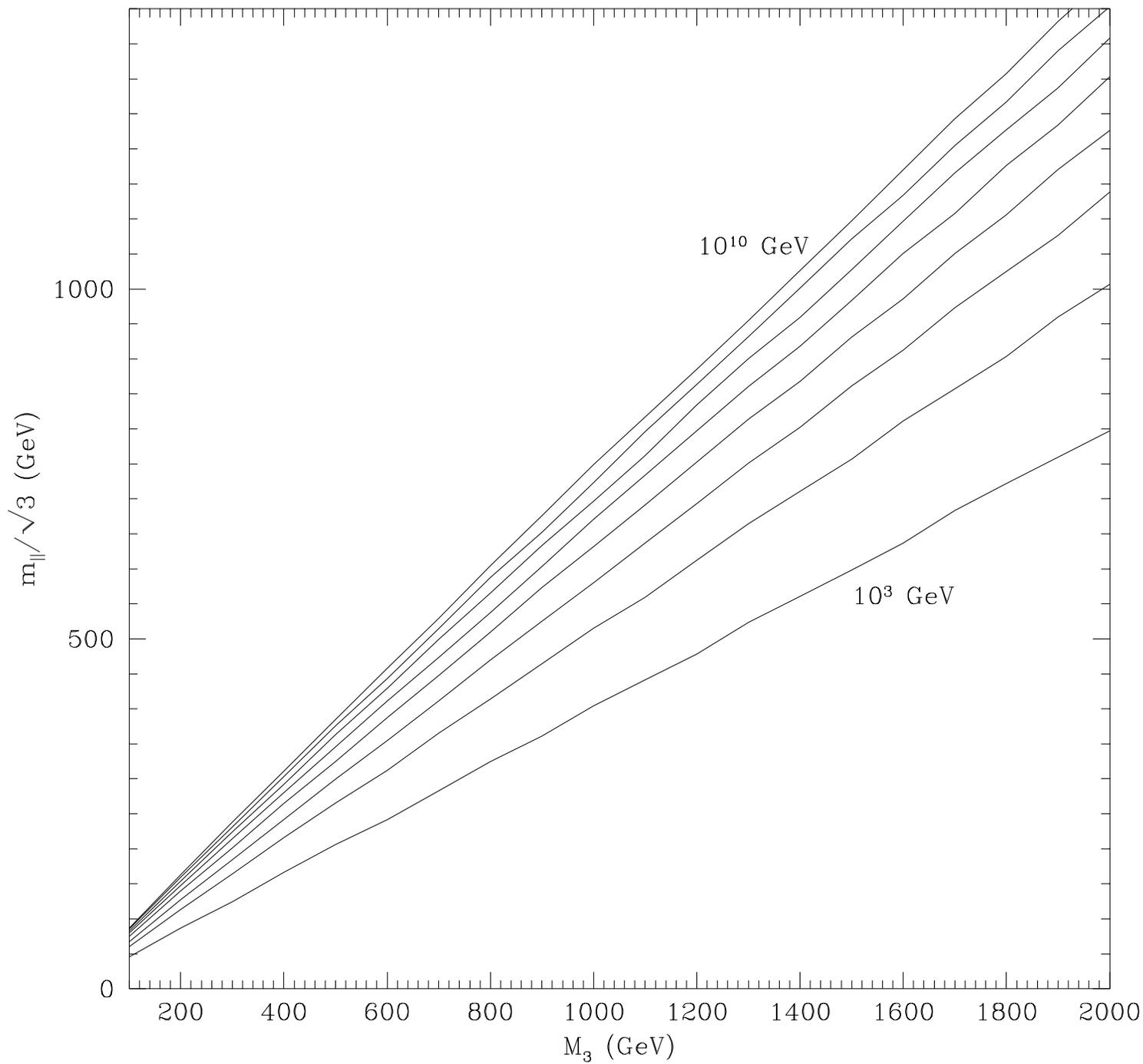
\end{document}